\documentclass[runningheads]{llncs}
\usepackage[T1]{fontenc}
\usepackage{newtxtext}       %
\usepackage[varvw]{newtxmath}       
\usepackage{array}

\usepackage{graphicx}
\usepackage{color}
\begin{document}
\title{Efficient high-order Discontinuous Galerkin simulations of multi-species high-enthalpy flows}
\titlerunning{Efficient high-order DG simulations of high-enthalpy flows}
%
\author{Beatriz Fagá\inst{1} \and
Georgii Oblapenko\inst{1,*} \and
Yiyang Zhu\inst{1,2} \and
Daniel Doehring\inst{1,3} \and
Manuel Torrilhon\inst{1}
}
\authorrunning{B. Fagá et al.}

%
\institute{Applied and Computational Mathematics, RWTH Aachen, 52062 Aachen, Germany \and
ETH Zürich, 8092 Zürich, Switzerland \and
Polaris Raumflugzeuge Gmbh, 28307 Bremen, Germany \\
Corresponding author: Georgii Oblapenko \email{oblapenko@acom.rwth-aachen.de}.}
\maketitle              
\begin{abstract}
Numerical simulation of hypersonic vehicles is an attractive approach to investigate the complex phenomena involved in such technologies. For this, second-order finite volume methods (FVM) are the established method in computational fluid dynamics, with the cost of high dissipation in affordable meshes, thus needing refined ones or adaptive mesh refinements techniques to obtain accurate results. An alternative is to make use of high-order discontinuous Galerkin methods, in particular discontinuous Galerkin spectral element method (DGSEM), once it achieves higher formal order of accuracy on element-local polynomial bases. Therefore the goal of this work is to present the numerical simulations of a two-dimensional scramjet configuration to test and verify the applicability of the DGSEM approach for a complex hypersonic flow where shock-shock interactions occur. Results revealed to be in agreement with reference FVM solvers, thus verifying the code implementation.


\end{abstract}
\section{Introduction}


Atmospheric flight above $\mathrm{Ma}=5$ remains one of the most challenging regimes in fluid mechanics. Air-breathing propulsion is attractive there because, unlike rockets, the vehicle does not need to carry its own oxidizer; but conventional ramjets become inefficient once decelerating the captured flow to subsonic speeds incurs large total-pressure losses and extreme stagnation temperatures~\cite{heiser1994}. The supersonic combustion ramjet, or scramjet, avoids this by keeping the flow supersonic throughout the intake, isolator, combustor and nozzle. Demonstrators such as HyShot~\cite{smart2006hyshot}, X-43A~\cite{marshall2005x43a} and X-51A~\cite{hank2008x51a} have shown the concept to be viable, yet design remains difficult and expensive: ground facilities cannot reproduce full flight enthalpies for more than milliseconds, leaving numerical simulation as the main alternative.
 
Nevertheless, the flow physics involved in such technology is hard to implement from the numerical point of view. A scramjet duct contains oblique shock trains, expansion fans, shock--shock and shock--boundary-layer interactions, high-gradients, and finite-rate chemistry coupled to thermal non-equilibrium. Therefore, a solver must be robust in the presence of strong discontinuities and accurate enough in smooth regions so that the small disturbances governing mixing are not taken away by numerical dissipation.

Second-order finite volume (FV) methods, combined with Riemann solvers and slope limiting, are the established approach in computational fluid dynamics: conservative, monotonicity-preserving and robust, at the cost of substantial dissipation on affordable meshes. High-order discontinuous Galerkin (DG) methods, in particular the discontinuous Galerkin spectral element method (DGSEM), instead achieve high formal order of accuracy on element-local polynomial bases, with a compact stencil that supports parallel implementation in high-perfomance computers~\cite{DURRWACHTER2021105039}, and with provable entropy stability. Their weakness is the Gibbs oscillations produced at steep gradients and shocks, which motivates shock-capturing strategies such as blending the DG operator with a subcell FV scheme~\cite{hennemann2021}, and entropy-conservative~\cite{tadmor2003entropy} schemes, which have been shown to improve stability in complex under-resolved simulations~\cite{chan2022entropy}. However, the implementation of realistic physics, such as multi-species flows, realistic thermodynamics, non-ideal gas laws, etc., in entropy-conservative DGSEM approaches is still ongoing, as it requires deriving and verifying new numerical schemes. 

In the present work, we use a two-dimensional scramjet configuration to test and verify the applicability and accuracy of the DGSEM approach for complex hypersonic flow scenarios with multiple shock-shock interactions, both for the case of a calorically perfect single-species gas, and the case of a multi-species flow with real thermodynamical data. The flow is simulated in \texttt{Trixi.jl}, an open-source Julia framework for adaptive high-order simulation of hyperbolic conservation laws~\cite{trixi}, using both its DGSEM discretisation and its subcell FV blending. The results are then assessed against the \texttt{rhoCentralFoam} solver of \texttt{OpenFOAM}~\cite{openfoam2024} and the \texttt{hyStrath} library developed at the University of Strathclyde~\cite{hystrath}, which extends \texttt{OpenFOAM} with solvers tailored to hypersonic and rarefied gas dynamics and thus provides an independently developed, widely used second-order FV reference. The comparison addresses shock structure and position as well as global flow quantities.

\section{Governing Equations}
We consider a two-dimensional flow of an inviscid compressible mixture of $N_c$ gases, described by the multi-component Euler equations:
\begin{equation}
    \frac{\partial}{\partial t}\mathbf{u} +  \frac{\partial}{\partial x}\mathbf{f}_x + \frac{\partial}{\partial y}\mathbf{f}_y= 0.\label{eqns:euler-generic}
\end{equation}
Here $t$ denotes the time, $x$ and $y$ are the spatial coordinates, $\mathbf{u}$ is the vector of conservative variables, and $\mathbf{f}_x$, $\mathbf{f}_y$ are the inviscid fluxes. Due to the moderate flow temperatures in the current setup ($T$ not exceeding 1100~K), we neglect chemical reaction terms in the above equations.
The vector of conservative variables  $\mathbf{u} \in \mathbb{R}^{N_c+3}$ is given by
\begin{equation}
     \mathbf{u} = \left(\rho_1, \ldots, \rho_{N_c}, \rho v_x, \rho v_y, E \right)^{\mathrm{T}},
\end{equation}
where $\rho_c$ is the density of species $c$, $v_x$ and $v_y$ are the flow velocities in the $x$ and $y$ directions, and $E=\rho e=\rho \varepsilon^{\mathrm{int}} + \rho v^2 / 2$ is the total flow energy, with the total flow density $\rho$ defined as a sum of the densities of the constituent flow species: $\rho=\sum_c \rho_c$. Here, $e$ is the specific energy, and $\varepsilon^{\mathrm{int}}$ is the specific internal energy computed as
\begin{equation}
    \varepsilon^{\mathrm{int}} = \sum_c Y_c \varepsilon^{\mathrm{int}}_c,
\end{equation}
where $Y_c=\rho_c / \rho$ and $\varepsilon^{\mathrm{int}}_c$ are given by
\begin{equation}
    \varepsilon^{\mathrm{int}}_c = \frac{3}{2}kT + \varepsilon^{\mathrm{f}}_c,\quad
    \varepsilon^{\mathrm{int}}_c = \frac{3}{2}kT +\varepsilon^{\mathrm{rot}}_c(T) + \varepsilon^{\mathrm{vibr}}_c(T) + \varepsilon^{\mathrm{f}}_c
\end{equation}
for atoms and molecules, respectively.

The inviscid fluxes are given by
\begin{equation}
     \mathbf{f}_x = \left(\rho_1 v_x, \ldots, \rho_{N_c} v_x, \rho v_x^2 + p, \rho v_x v_y, (E+p)v_x \right)^{\mathrm{T}},
\end{equation}
\begin{equation}
     \mathbf{f}_y = \left(\rho_1 v_y, \ldots, \rho_{N_c} v_y, \rho v_x v_y, \rho v_y^2 + p, (E+p)v_y \right)^{\mathrm{T}}.
\end{equation}
where $p$ is the pressure, given by the ideal gas law: $p=nkT$, where $n$ is the number density, $k$ is the Boltzmann constant, and $T$ is the flow temperature.

\section{Numerical Method}
For brevity, we focus on this section only on the evaluation of entropy-conservative numerical fluxes for the multi-species Euler equations, without going into detail about the underlying discretization.

Defining the integral
\begin{equation}
    \eta(T)_c := \int_0^T \frac{c_{v,c}(\tau)}{\tau}\mathrm{d}\tau,\label{eq:entropy-integral-def}
\end{equation}
we can write the physical entropy $s$ as
\begin{equation}
    s = \sum_c Y_c \eta_c - \sum_c Y_c \frac{k}{m_c} \ln \rho_c = \sum_c Y_c \eta_c(T) -  \sum_c Y_c \frac{k}{m_c} \ln \rho_c. \label{eq:entropy-def}
\end{equation}
This serves as a basis for constructing entropy-conservative fluxes for gases with internal energies. In the present work, we use the approach proposed in~\cite{oblapenko2025entropy} for single-species flows and extended in~\cite{oblapenko2026entropy} to the multi-species case.
The approach works by using piece-wise polynomial interpolations for the specific heats $c_{v,c}$ and internal energies $\varepsilon_c^{\mathrm{int}}$, and using exact integration in~\eqref{eq:entropy-integral-def}. Therefore, while this approach induces a \textit{modelling} error (due to the discretization and approximation of the thermodynamic quantities), the entropy-conservative fluxes remain \textit{exact} for the given model.
In the present work, we use a piece-wise linear interpolation with a step-size $\Delta T$ of $1$~K.
Denoting by $c_{v,c,i}$ the stored value of $c_{v,c}(T)$ computed at a tabulated temperature of $T_i$, and defining $N$ as $\lfloor (T - T_{\mathrm{min}}) / \Delta T \rfloor$, where $T_{min}$ is the minimum temperature used for the tabulation, we can define partial values of the integral part of the entropy:
\begin{equation}
    \eta_{c,i} := \left(c_{v,c,i} - \frac{(c_{v,c,i+1} - c_{v,c,i}) T_{i}}{\Delta T}\right) \ln\left(\frac{T_{i+1}}{T_{i}}\right) + \left(c_{v,c,i+1} - c_{v,c,i}\right).\label{eq:eta_tables}
\end{equation}
This leads to the following expression for $\eta_c(T)$:
\begin{equation}
  \eta_c(T) = \sum_{i=0}^{N-1} \eta_{c,i} + \left(c_{v,c,N} - \frac{(c_{v,c}(T) - c_{v,c,N}) T_{N}}{\Delta T}\right) \ln\left(\frac{T}{T_{N}}\right) + \left(c_{v,c}(T) - c_{v,c,N}\right).\label{eq:eta_compute}
\end{equation}
In the solver, the sums $\sum_{i=0}^{N-1} \eta_{c,i}$ are precomputed and stored for different values of $N$ at the start of the simulation, along with the tabulated values of the thermodynamical data.

Introducing brackets to the jump of a variable between two points:  $\llbracket a \rrbracket$ = $a_+ - a_-$, we can formally define
\begin{equation}
  \frac{c_{v,c}\left(T^\ast\right)}{T^\ast} := \frac{\left \llbracket \eta_c(T) \right \rrbracket}{\llbracket T \rrbracket},\quad
  c_{v,c}\left(T^{\ast\ast}\right) := \frac{\left \llbracket \varepsilon^{\mathrm{int}}_c \right \rrbracket}{\llbracket T \rrbracket }.\label{eq:jump-eint}
\end{equation}
Finally, introducing averaging operators~\cite{ranocha2018comparison}
\begin{equation}
  \left\{\!\left\{ a \right\}\!\right\} = \frac{1}{2}\left(a_- + a_+\right),\quad
  \left\{\!\left\{ a \right\}\!\right\}_{\mathrm{geo}} = \sqrt{a_- a_+},\quad
  \left\{\!\left\{ a \right\}\!\right\}_{\mathrm{log}} = \frac{a_+ - a_-}{\log a_+ - \log a_-},
\end{equation}
we can write expressions for the components of the entropy-conservative flux in the $x$-direction (we note that the flux in the $y$-direction is computed similarly, and the flux in a general direction is a linear combination of the directional fluxes):
\begin{eqnarray}
  F^{\mathrm{num},x}_{\rho_c} = & {\left\{\!\left\{ \rho_c \right\}\!\right\}_{\mathrm{log}}}\left\{\!\left\{ v_x \right\}\!\right\},\:c=1,\ldots,N_c,\label{eq:flux_fx_rho} \\
  F^{\mathrm{num},x}_{\rho v_x}  = & \left\{\!\left\{ v_x \right\}\!\right\} \sum_c F^{\mathrm{num},x}_{\rho_c}   + \sum_c\frac{k}{m_c}\frac{ \left\{\!\left\{ \rho_c \right\}\!\right\}}{ \left\{\!\left\{ 1/T \right\}\!\right\}}, \\
  F^{\mathrm{num},x}_{\rho v_y}  =& \left\{\!\left\{ v_y \right\}\!\right\} \sum_c F^{\mathrm{num},x}_{\rho_c} , \\
  F^{\mathrm{num},x}_{E} = & \sum_c F^{\mathrm{num},x}_{\rho,c} \left(
  \left\{\!\left\{ T \right\}\!\right\}_{\mathrm{geo}}^2 \left(\frac{c_{v,c}\left(T^\ast\right)}{T^\ast} -\left\{\!\left\{ \frac{1}{T} \right\}\!\right\} c_{v,c}\left(T^{\ast\ast}\right) \right) \nonumber \right.  \\
  & + \left. \left\{\!\left\{ \varepsilon^{\mathrm{int}}_c \right\}\!\right\} - \frac{\left\{\!\left\{ v_x^2 \right\}\!\right\} +\left\{\!\left\{ v_y^2 \right\}\!\right\}} {2}\right) + \left\{\!\left\{ v_x \right\}\!\right\} F^{\mathrm{num},x}_{\rho v_x} +  \left\{\!\left\{ v_y \right\}\!\right\}  F^{\mathrm{num},x}_{\rho v_y} .\label{eq:flux-E_x}
\end{eqnarray}
Thus, to compute the fluxes, one first computes the primitive variables $(\rho_1,\ldots,\rho_{N_c},v_x,v_y,T)$ from the conservative variables $\mathbf{u}$ via a Newton iteration. The computed value of temperature is then used to calculate the specific heats and internal energies, the former also required for computation of the sound speed for wave-speed estimation in the local Lax-Friedrichs dissipation term used across cell interfaces. Next, the integral part of the entropy is computed via~\eqref{eq:eta_compute}, and the quantities defined in~\eqref{eq:jump-eint} are calculated.
After this, the entropy-conservative flux is computed. In the DGSEM approach used in the present work, the entropy-conservative flux is used to compute the volume flux terms, and a local Lax-Friedrichs dissipative flux is used across cell interfaces.

\subsection{Improving Computational Efficiency of DG Methods}
A significant part of the computational costs in entropy-conservative DGSEM methods comes from the volume flux evaluations, due to both the flux differencing approach require more flux evaluations~\cite{doehring2026volume} and the fact that entropy-conservative fluxes oftentimes contain averaging operators that make use of logarithmic functions~\cite{ranocha2018comparison}, which have a higher computational cost than more simple arithmetic or geometric averaging.
In the approach presented above, additional logarithm evaluations are required in~\eqref{eq:eta_compute} (the values $\eta_{c,i}$ defined by ~\eqref{eq:eta_tables} are pre-computed at the start of the simulation).
However, by noting that $T_N$ is the closest tabulated temperature value lesser or equal to $T$, we can write the ratio $T/T_{N}$ as
\begin{equation}
    \frac{T}{T_N} = 1 + \frac{\delta T}{T_N}, \delta T = T - T_N \geq 0.
\end{equation}
If we assume that the chosen temperature discretization step $\Delta T$ used for the tabulation of the thermodynamic quantities is smaller than $T_{\mathrm{min}}$, which is reasonable, given that for temperatures below 50-60~K, gases would liquefy and the flow equations would lose their validity, then we can assume that
\begin{equation}
    \frac{\delta T}{T_N} \leq \frac{\Delta T}{T_{\mathrm{min}}} < 1.
\end{equation}
This allows us to perform a Taylor series expansion of the logarithm appearing in~\eqref{eq:eta_compute}:
\begin{equation}
    \ln\left(\frac{T}{T_{N}}\right) = \frac{\delta T}{T_N} - \frac{1}{2}\left(\frac{\delta T}{T_N}\right)^2 + \frac{1}{3}\left(\frac{\delta T}{T_N}\right)^3 + \mathcal{O}\left(\left(\frac{\delta T}{T_N}\right)^4\right).\label{eq:taylor-approx}
\end{equation}
Thus, retaining only the terms up to the third-order, we can expect a computational efficiency gain in the volume flux evaluation, whilst retaining almost-exact entropy conservation.

\section{Numerical Simulation Results}
We performed numerical simulations of a hypersonic inviscid non-reacting flow in a scramjet intake at Mach 7.3. We consider both a single-species calorically perfect gas to verify the correctness robustness of the DGSEM-based simulations, and a more realistic test case with a two-species non-reactive mixture of oxygen and nitrogen, with correct thermodynamical data.

\begin{figure}[h!]
    \centering
    \includegraphics[width=0.85\linewidth]{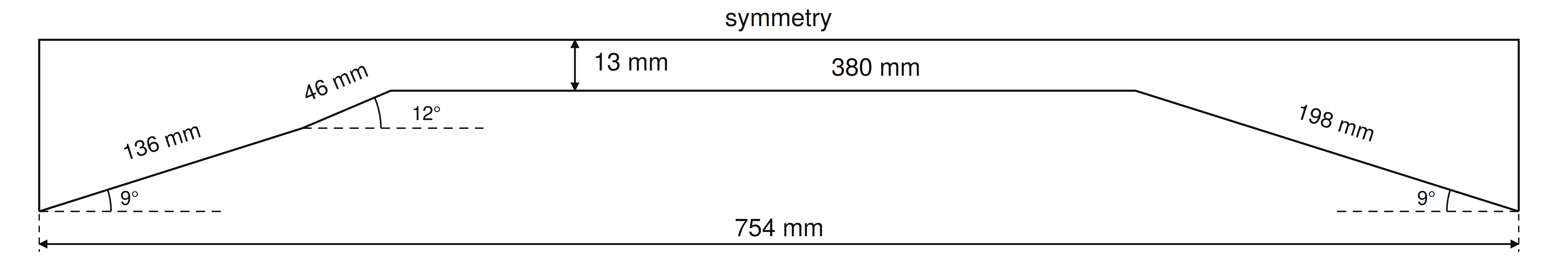}
    \caption{Domain dimensions. Adapted from~\cite{hoste2017numerical}.}
    \label{fig:geometry}
\end{figure}

The case is composed of a two ramp compression inlet followed by a combustor and expansion region, as shown on~Figure \ref{fig:geometry}. A 2-D axisymmetric half-domain was meshed using Gmsh~\cite{gmsh2009} with only quadrilateral elements, and to assess that numerical results are not significantly affected by spatial discretizations, we performed a mesh independence analysis. 
To this end, we performed simulations using the multi-species gas model in \texttt{AeroTrixi.jl} on three meshes with 316035 (fine), 157130 (medium), and 79628 (coarse) elements. The discretization uncertainty was estimated with the Grid Convergence Index (GCI)~\cite{Celik2008}, using the temperature at the combustor entrance (x = 0,17932 $m$) and exit (x = 0,55932 $m$).
The average observed order of convergence $p_{ave}$ lies between 1,63 and 3,49, with the uncertainty of the fine mesh in relation to a Richardson extrapolation below 0,2\% ($\mathrm{GCI}_\text{fine}^{21}$). For the purpose of obtaining more accurate results, the fine mesh was selected for the simulations. 

With the fine mesh, a free-stream velocity of 2800~$\mathrm{m}$ $\mathrm{s^{-1}}$, pressure of 4100~$\mathrm{Pa}$ and temperature of 370~$\mathrm{K}$ is applied to the inlet. A zero-heat-flux is applied to the lower and upper walls, thereby neglecting heat transfer on these boundaries, while a supersonic outflow boundary condition is used at the exit to extrapolate the boundary flux from the interior.

The DG simulations were carried using second-order polynomials, i.e. $p=2$. A high-order polynomial representation oscillates across a discontinuity, so a shock-capturing mechanism is necessary. The approach of~\cite{hennemann2021}, where the indicator estimates the local smoothness of the solution and maps onto a blending factor ($\alpha \in [\alpha_{\min}, \alpha_{\max}]$, where $\alpha_{\min} = 0.01$ and $\alpha_{\max} = 0.3$). A first-order finite volume scheme using the local Lax-Friedrichs flux with a maximum absolute wave speed estimator is then blended with the high-order DG scheme to ensure damping of oscillations.
\texttt{Trixi.jl} and the extension package \texttt{AeroTrixi.jl} were applied to perform the simulations; the setups to reproduce the results are available in a dedicated reproducibility repository~\cite{reprorepo}.


\subsection{Calorically Perfect Gas}

The inviscid, single-species model was implemented in both \texttt{Trixi.jl} and \texttt{OpenFOAM}. The same boundary and initial conditions described are applied in \texttt{OpenFOAM}; specifics of the setup are described in \ref{tab:solver_settings}. Here $\gamma$ refers to the adiabatic index of the gas.

\begin{table}[htbp]
  \centering
  \small
  \caption{Solver configurations: \texttt{rhoCentralFoam} (\texttt{OpenFOAM}) and \texttt{Trixi.jl}.}
  \label{tab:solver_settings}
  \begin{tabular}{@{}ccc@{}}
    \hline
    Setting & rhoCentralFoam & \texttt{Trixi.jl} \\
    \hline
    Version                   & \texttt{OpenFOAM} v2412 & \texttt{Trixi.jl} 0.16.3 \\
    Equations / model         & Single species, $\gamma = 1.4$, $\mu = 0$ & Compressible Euler, $\gamma = 1.4$ \\
    Mesh                      & Gmsh & P4estMesh \\
    Discretization            & Finite Volume & DG, polynomial degree $N = 2$ \\
    Convective/volume flux    & Kurganov & Ranocha \\
    Surface flux              & -- & Lax Friedrichs \\
    Limiter / shock indicator & vanLeer (baseline) & Hennemann and Gassner \\
    Time integration          & Backward Euler, adjustable $\Delta t$ & SSPRK53, CFL-limited \\
    Maximum Courant number    & 0.1 & 2.5 \\
    \hline
  \end{tabular}
\end{table}
\begin{figure}[h!]
    \centering
    \includegraphics[width=0.9\linewidth]{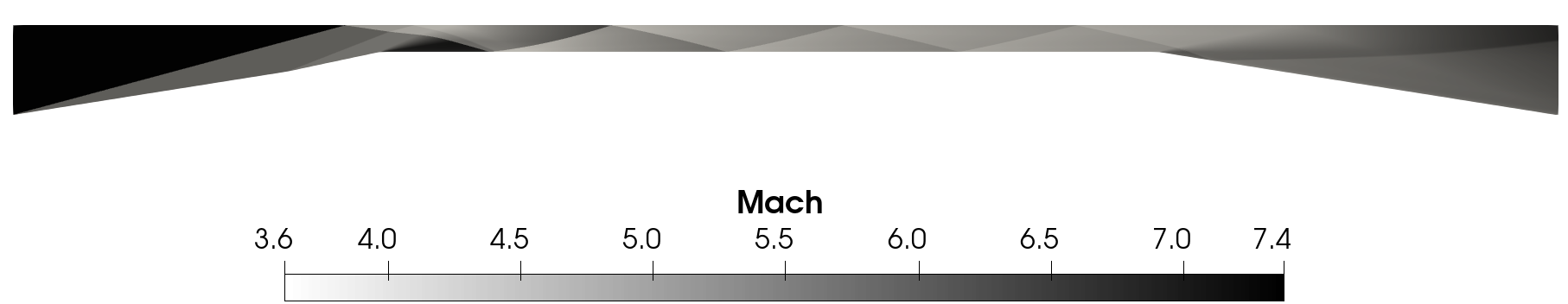}
    \caption{Mach number distribution for calorically perfect simulation with \texttt{Trixi.jl}.}
    \label{fig:machdistr}
\end{figure}
\begin{figure}[h!]
    \centering
    \caption{Pressure distribution profile at 1 mm from bottom-wall for calorically perfect simulations: \texttt{OpeFOAM} vs \texttt{Trixi.jl}.}
    \includegraphics[width=0.85\linewidth]{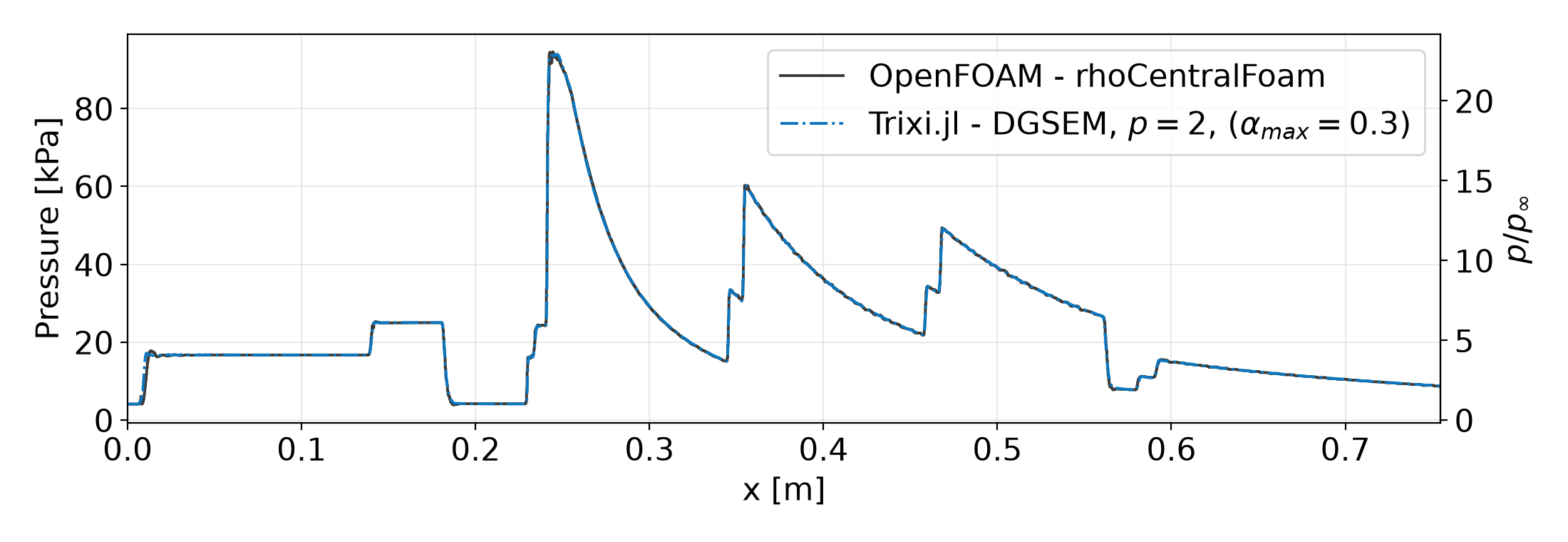}
    \label{fig:bottomwallp}
\end{figure}
Figure~\ref{fig:machdistr} shows the distribution of the Mach number in the flowfield for the simulation performed with \texttt{Trixi.jl}. The result shows expected behaviour of the shock structures, consistent both with previous numerical studies~\cite{hoste2017numerical} and the original experiments~\cite{lorrain2012}.

Furthermore, Figure~\ref{fig:bottomwallp} shows the distribution of the pressure along the bottom wall, computed with \texttt{OpeFOAM} and \texttt{Trixi.jl}. We observe excellent agreement throughout the whole domain, thus verifying our implementation of the meshing procedure and boundary conditions.

\subsection{Multi-species Gas Model}
We now consider a multi-species gas flow with realistic thermodynamics. We model a mixture of N$_2$ and O$_2$, assuming continuous rotational spectra and the harmonic oscillator model for the vibrational spectra, with characteristic vibrational temperatures of 3371 $K$ (N$_2$) and 2273.52 $K$ (O$_2$). Due to the moderate temperatures, we neglect the effects of electronic excitation. The molar fraction of nitrogen and oxygen in the inflow are assumed to be 0.79\% and 0.21\%, respectively. The case setup is shown in Table \ref{tab:multispeciesmodel}.

\begin{table}[htbp]
  \centering
  \small
  \caption{Solver configurations: \texttt{hy2Foam} (\texttt{hyStrath}) and \texttt{AeroTrixi.jl}, two-species (N$_2$/O$_2$) scramjet.}
  \label{tab:multispeciesmodel}
  \begin{tabular}{@{}ccc@{}}
    \hline
    Setting & \texttt{hy2Foam} & \texttt{AeroTrixi.jl} \\
    \hline
    Version                   & \texttt{OpenFOAM} v1706 + \texttt{hyStrath} & \texttt{Trixi.jl} 0.4.0 \\
    Equations / model         & thermally perfect & Multi-species compressible Euler \\
    Mesh                      & \texttt{Gmsh} & \texttt{P4estMesh}  \\
    Discretization            & Finite Volume & DG, polynomial degree $N = 2$ \\
    Convective/volume flux    & Kurganov-Noelle-Petrova& Oblapenko \\
    Surface flux              & -- & Lax Friedrichs \\
    Limiter / shock indicator & vanLeer & Hennemann and Gassner \\
    Time integration          & Backward Euler, adjustable $\Delta t$ & SSPRK53, CFL-limited \\
    Courant / CFL number      & 0.1 (maximum) & 2.5 \\
    \hline
  \end{tabular}
\end{table}

\begin{figure}[h!]
    \centering
    \caption{Pressure (top) and temperature (bottom) profile at 1 mm from bottom-wall boundary.}
    \includegraphics[width=0.85\linewidth]{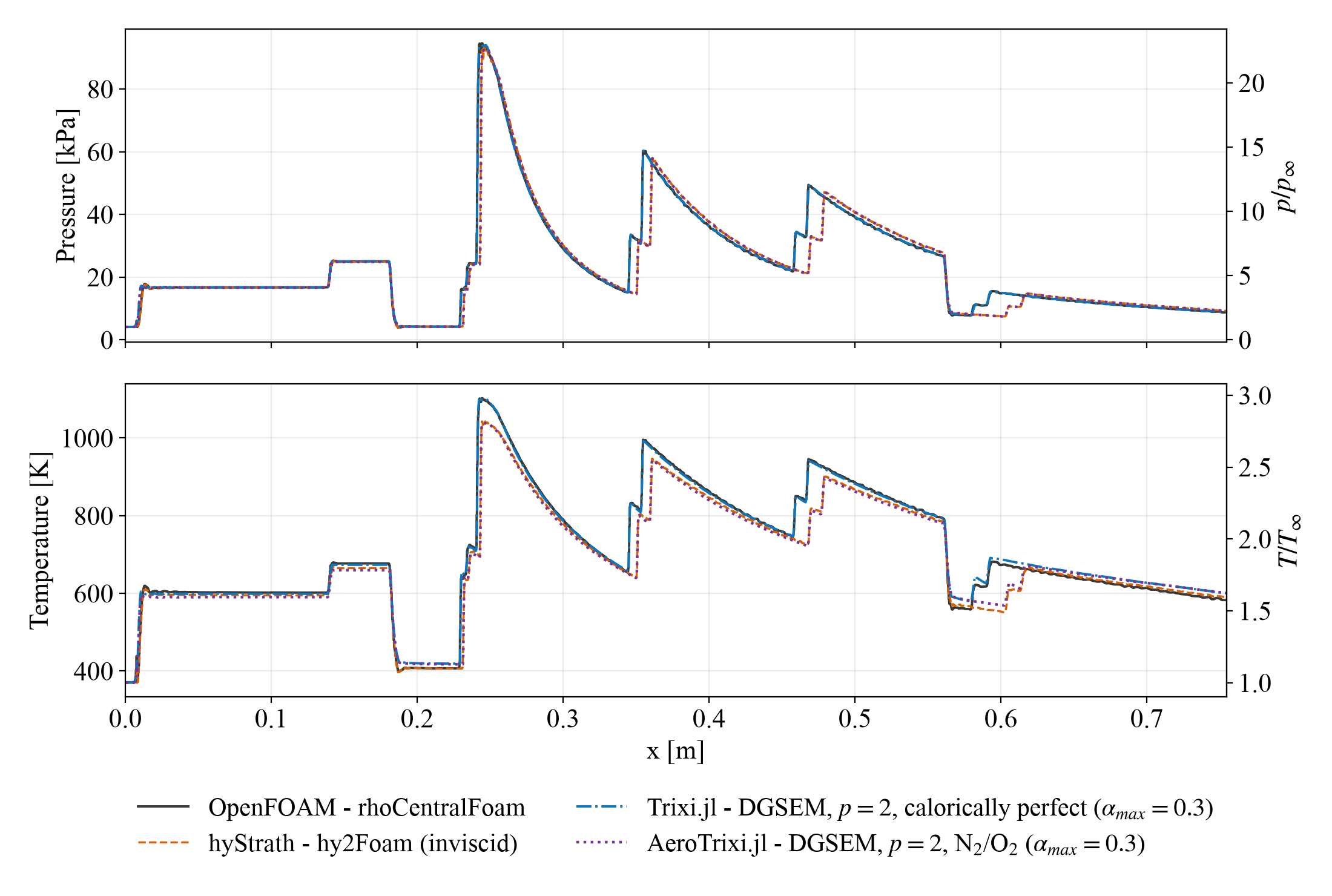}
    \label{fig:pressure_temperature_all}
\end{figure}

For comparison, the case is assessed against the \texttt{hy2Foam} solver from the \texttt{hyStrath} library~\cite{hystrath}, a well validated open-source code for atmospheric entry analysis. Figure \ref{fig:pressure_temperature_all} presents the pressure and temperature distribution at 1 .0 mm from the bottom-wall boundary. for both cases, with the results obtained using the calorically perfect model (\texttt{OpenFOAM} and \texttt{Trixi.jl}) and the two-species gas model with real thermodynamical data (\texttt{hyStrath}, \texttt{AeroTrixi.jl}). 

It can be seen that within each gas model, both solvers agree remarkably well, with two important notes, at the inlet and the downstream the domain. First, for both \texttt{OpenFOAM} based solvers, its possible to observe small peaks at shock incidence at the leading edge; this isn't physical, but rather numerical . Since all cases are inviscid, the wall pressure is a step at every corner, thus, this is likely caused by the insufficient numerical dissipiation in the existing~\texttt{OpenFOAM} setup, whereas the use of subcell shock-capturing blending with the prescribed $\alpha_{\max}$ adds enough numerical dissipation at shocks and corners, sufficiently damping these extrema. 

\begin{figure}[h!]
    \centering
    \caption{Pressure profiles across the first (left) and last (right) shocks in the scramjet.}
    \includegraphics[width=0.85\linewidth]{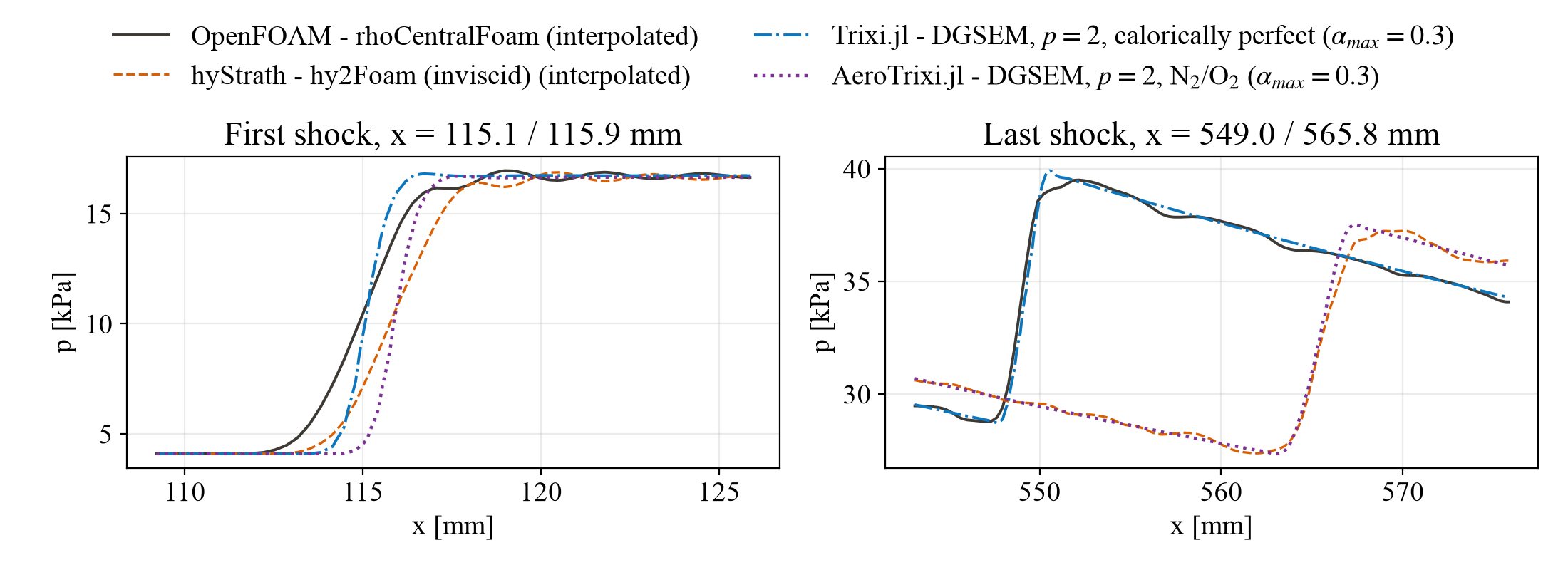}
    \label{fig:shocks}
\end{figure}

Second, in the freestream region the vibrational modes of N$_2$ and O$_2$ are practically frozen, so both models behave similarly. Behind the successive shocks, the two-species model lies progressively further downstream because of the rise in temperature at the shock formation (Figure \ref{fig:pressure_temperature_all}), and since part of the compression energy is stored in vibration, $\gamma$ and temperature drop after the shock, affecting the entire flow, which consequently accumulates the profile shift along the shock train. To study the individual shocks in more detail, we consider the pressure profiles across the first and last shocks, as seen on Figure~\ref{fig:shocks}. Small discrepancies are visible, which is attributed to the choice of flux limiters and the discretization method itself. Due to the higher order of the DG simulation, and a more detailed shock-capturing scheme, the first strong shock (as seen on the left subplot) is sharper in the DG simulation. The weaker shocks at the right end of the scramjet are more similar in both the finite volume and DG simulations, as the pressure gradients are smaller in magnitude. Nevertheless, the results show excellent agreement in shock formation, further supporting the verification of the implementation.

Finally, we performed a study regarding the computational efficiency of the two-species simulation in \texttt{Trixi.jl}. Using a bisection method, we determined the maximum allowable CFL number within the SSPRK53 method to be 3.25. Using the Taylor series-approximation~\eqref{eq:taylor-approx} in \eqref{eq:eta_compute} leads to an approximately 15\% reduction in the required computational time, with the results of the simulations being virtually unchanged. For higher-order DGSEM simulations we expect the speed-up to be even larger, due to the increasing cost of evaluating the volume fluxes.

\section{Conclusions}
We have studied the performance of entropy-conservative higher-order DGSEM simulations applied to supersonic simulations of a scramjet intake. We first verified the solver for the case of a calorically perfect gas against a reference implementation in \texttt{OpenFOAM}. For a more realistic of a two-species flow with real thermodynamical data, we assessed the impact of using an approximate version of a previously developed entropy-conservative flux, verifying the implementation by also comparing to \texttt{hy2Foam}, a reference solver dedicated to complex atmospheric reentry flows.
Future work will incorporate viscous effects for a more realistic benchmarking and cross-verification of the developed DGSEM solver.

\begin{credits}
\subsubsection{\ackname} This work has been supported by the German Research Foundation within the research unit DFG-FOR5409.

\subsubsection{\discintname}
The authors have no competing interests to declare that are
relevant to the content of this article.
\end{credits}
%
%
%


\end{document}